\documentclass[]{iopart}
\usepackage{iopams}
\usepackage[]{graphicx}
\usepackage{color}

\newcommand{\dd}{d_{220}}
\newcommand{\eSi}{{\rm ^{28}Si}}
\newcommand{\NA}{N_{\rm A}}

\begin{document}

\title[]{Comparison of the INRIM and PTB lattice-spacing standards}
\author{E Massa\ddag, G Mana\ddag\; and U Kuetgens\dag}
\address{\ddag INRIM - Istituto Nazionale di Ricerca Metrologica, Str.\ delle Cacce 91, 10135 Torino, Italy}
\address{\dag  PTB - Physikalish-Technische Bundesanstalt, Bundesalle 100, 38116 Braunschweig, Germany}
\ead{e.massa@inrim.it}

\begin{abstract}
To base the kilogram definition on the atomic mass of the $\eSi$ atom, the present relative uncertainty of the $\eSi$ lattice parameter must lowered to $3\times 10^{-9}$. To achieve this goal, a new experimental apparatus capable of a centimetre measurement-baseline has been made at the INRIM. The comparison between the determinations of the lattice parameter of crystals MO*4 of INRIM and WASO 4.2a of PTB is intended to verify the measurement capabilities and to assess the limits of this experiment.
\end{abstract}

\submitto{Metrologia}
% 06.20.-f metrology
% 06.20.Jr fundamental constantsù
% 61.05.C- x-ray diffraction
\maketitle

%\large
%\baselineskip 9mm

\section{Introduction}
To determine the Avogadro constant, $\NA$, to an accuracy allowing the kilogram definition to be based on the atomic mass of the $\eSi$ atom \cite{kg,NA}, several metrology institutes participate in a research project (International Avogadro Coordination, IAC) for the determination of $\NA$ with the use of a highly enriched $\eSi$ crystal. Isotope enrichment and crystal production were completed and a 5 kg crystal with $\eSi$ enrichment higher than 99.99\% is now available for measurements \cite{PB-28Si}.

In this framework, the relative uncertainty of the (220) Si lattice-plane spacing measurement by combined x-ray and optical interferometry must be reduced to $3 \times 10^{-9}$. With this in view, INRIM developed a guide capable of interferometer displacements up to 5 cm with guiding errors commensurate with the requirements of atomic-scale positioning and alignment. Test measurements were started to verify the actual measurement capabilities and to assess experimental limits. In the first place, the MO*4 crystal of INRIM \cite{d220-PRL} was integrated in the apparatus and a $6\times 10^{-9}$ relative measurement accuracy was demonstrated \cite{d220-OEXP}. This exercise was repeated with the WASO 4.2A crystal of PTB \cite{d220-PTB}; the results are related in the present paper.

In addition to gain more confidence in the performances of the new set-up, measurement repetitions were prompted by the improved measurement capabilities. The high purity of the WASO 4.2A crystal made it a standard for many precision measurements \cite{CODATA}. Furthermore, in 1998, the MO*4 crystal was sent to the PTB for comparison with the WASO 4.2A crystal \cite{Martin}. Therefore, a relative lattice spacing measurement value is available for a consistency check.

\section{Experimental apparatus}
\begin{figure}
\centering
\includegraphics[width=80mm]{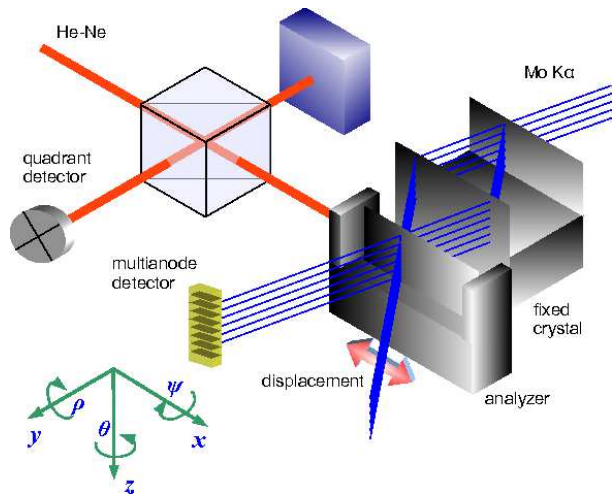}
\caption{Combined X-ray and optical interferometer.}\label{xint}
\end{figure}

\begin{figure}
\centering
\includegraphics[width=75mm]{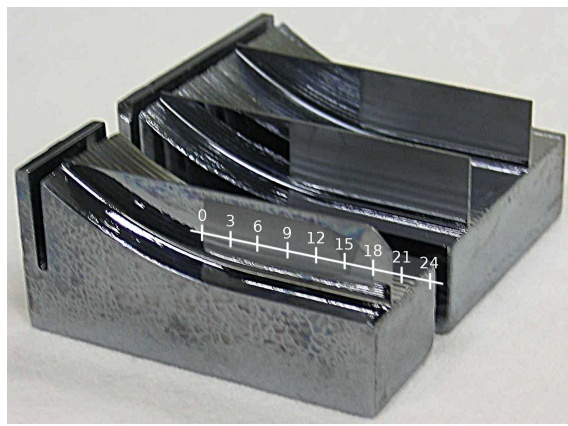}
\caption{Photograph of the WASO 4.2A interferometer. The ruler indicates where the $\dd$ measurements were carried out. The re-machined area is also visible.} \label{waso42a}
\end{figure}

The WASO 4.2A interferometer is shown in Figs.\ \ref{xint} and \ref{waso42a}. It consists of three flat and parallel crystals so cut that the (220) planes are orthogonal to the crystal surfaces. X rays from a 17 keV Mo K$_\alpha$ source, having a $(10\times 0.1)$ mm$^2$ line focus, are split by the first crystal and then recombined, via two transmission mirror crystals, by the third, called analyzer. When the analyzer crystal is moved in a direction orthogonal to the (220) planes, a periodic variation of the transmitted and diffracted x-ray intensities is observed, the period being the diffracting-plane spacing. The analyzer embeds a front mirror, so that the crystal displacement is measured by optical interferometry; the necessary picometer resolution is achieved by polarization encoding and phase modulation. According to the measurement equation
\begin{equation}\label{me}
 \dd = (m/n)\lambda/2,
\end{equation}
where $n$ is the number of x-ray fringes of $\dd$ period observed in a crystal displacement spanning $m$ optical fringes of $\lambda/2$ period. The successful operation of a separate-crystal interferometer is a challenge: the fixed and movable crystals must be so faced to allow the atoms to recover their exact position in the initial single crystal and they must be kept aligned notwithstanding the analyzer displacement.

With respect to all previous apparatuses, the key development is the extension of the interferometer operation to displacements of 5 cm. This measurement capability is obtained by means of a guide where an L shaped carriage slides on a quasi-optical rail. An active tripod with three piezoelectric legs rests on the carriage. Each leg expands vertically and shears in the two, $x$ and $y$, transverse directions, thus allowing compensation for the sliding errors and electronic positioning of the x-ray interferometer over six degrees of freedom to atomic-scale accuracy. Crystal displacement, parasitic rotations, and transverse motions are sensed via laser interferometry and by capacitive transducers. Feedback loops provide picometer positioning, nanoradian alignment, and interferometer movement with nanometer straightness.

The lattice spacing is determined by starting from the approximation $n/m \approx 1648.28$ and by measuring the x-ray fringe fractions at the displacement ends to an accuracy sufficient to predict the number of lattice planes in increasing analyzer displacements. To estimate the x-ray fringe fraction, the least squares method was applied; the input data were samples of about six fringes, about 300 samples with a 100 ms integration time and a sample duration of 30 s. Since it was not possible to keep the drift between the x-ray and optical interferometers as small as desirable, the analyzer was repeatedly moved back and forth along any given displacement. Each measurement is thus typically the average of about 9 values collected in measurement cycles lasting the half hour during which the analyzer is moved back and forth by about 1 mm (3000 optical orders or $5\times 10^6$ lattice planes).

\section{Measurement results}
\begin{figure}
\centering
\includegraphics[width=75mm]{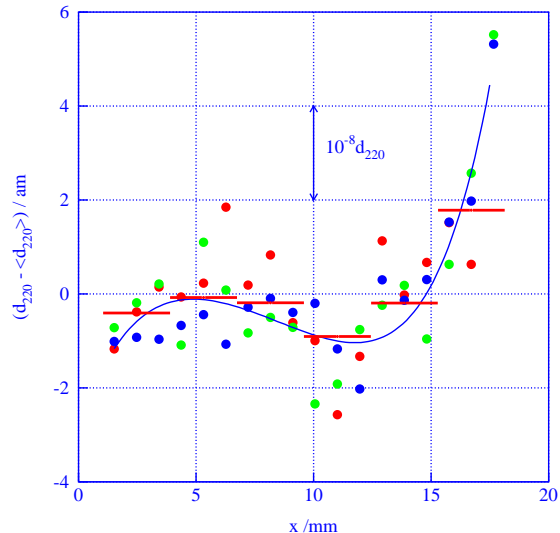}
\caption{Lattice spacing variation in the WASO 4.2A crystal. Measurements have been carried out on 2008-08-06 (red), 2008-09-16 (green), and 2008-10-15 (blue). Horizontal lines indicate the measurements over the crystal slices expressed by the bar length.}\label{map}
\end{figure}

Figure \ref{map} shows the results of three surveys of the lattice spacing values along the line indicated in Fig.\ \ref{waso42a} carried out from June to October 2008. During surveys, the analyzer was shifted step-by-step while the splitter/mirror crystal and x rays were maintained fixed. The measurements were carried out over 18 contiguous crystal slices, about 1 mm wide. Additional measurements were also carried out over 6 contiguous crystal slices, 3 mm wide; the consistency between the results of these measurements and the previous ones is shown in Fig.\ \ref{map}. Figure \ref{map} shows a significant rise of the measurement results at the crystal edge. The reason is a thickness gradient of the analyzer; after PTB measured the lattice spacing early in the 1980s, the geometrical perfection of the WASO 4.2A crystal was impaired, so that it was re-machined and, as shown Fig.\ \ref{waso42a}, a usable area of about $(18 \times 6)$ mm$^2$, 0.390 mm thick, was obtained. At the crystal edge this area vanishes and the relevant geometrical error causes a phase shift of the x-ray fringes which is translated into a lattice spacing variation.

After account was taken of the corrections and of the uncertainty contributions listed in Table I, the values shown in Fig.\ \ref{map} were averaged. The final average lattice-spacing value in a vacuum and at 22.5 $^\circ$C, is
\begin{equation}\label{d220value}
 \dd({\rm WASO 4.2A}) = 192.0155691(10)  \textrm{ pm}.
\end{equation}
For the reason mentioned, the last three data at the crystal edge were not included in the average. The value in (\ref{d220value}) is not corrected for carbon and oxygen contamination. Consequently, the relevant contribution to the error budget was not taken into account.

A detailed discussion of the corrections and of the uncertainty contributions can be found in \cite{d220-OEXP}; additional information is now given. The first concerns the correction for and the uncertainty due to imperfect crystal trajectory. Since the lattice spacing is measured by comparing the projections of the crystal displacement over the normals to the diffracting planes and the front mirror, when the projection angles are different an error occurs. This error is proportional to the angle between the front mirror and diffraction planes which angle, in the WASO 4.2A crystal, is quite small, 0.039 mrad. This explains the negligible correction and the very small uncertainty contribution.

\begin{table}
\renewcommand{\arraystretch}{1.3}
\caption{Crystal WASO 4.2A -- corrections and uncertainties of $d_{220}$.}
\begin{center} {%\tiny
\begin{tabular}{@{}lrr} \hline\hline
contribution &correction/$10^{-9}$ &uncertainty/$10^{-9}$\\
\hline
 statistic              & 0.0    & 1.3 \\
 wavelength             &$-3.0$  & 1.5 \\
 laser beam diffraction & 7.3    & 0.7 \\
 laser beam alignment   & 1.3    & 1.8 \\
 Abbe's error           & 0.0    & 2.5 \\
 trajectory             &0.1     & 0.3 \\
 temperature            &$-2.0$  & 3.0 \\
 aberrations            & 0.0    & 2.0 \\
\hline
 total                  & 3.7    & 5.2 \\
\hline\hline
\end{tabular} } \end{center} \end{table}

\begin{figure}
\centering
\includegraphics[width=75mm]{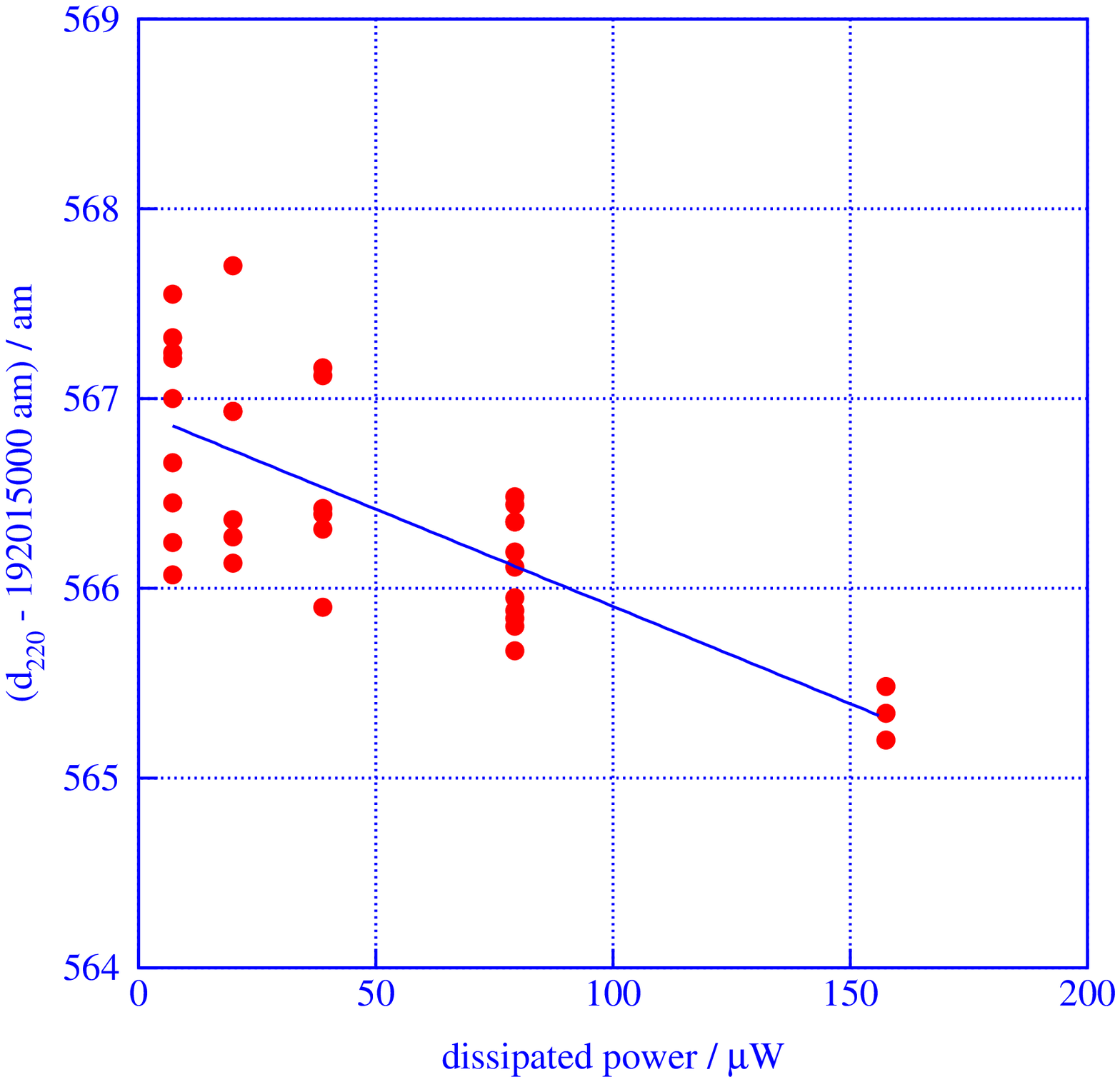}
\caption{Apparent variations of the lattice plane spacing when the crystal temperature is measured with different currents -- 0.27 mA, 0.45 mA, 0.62 mA, .89 mA, and 1.25 mA.}\label{self-heat}
\end{figure}

The crystal temperature was measured by a primary capsule 100 $\Omega$ Pt resistance-thermometer located on the silicon platform supporting the analyzer; the measurement current was 0.27 mA. The thermometer self-heating was investigated by repeating the lattice spacing measurement with different currents; the results are shown in Fig.\ \ref{self-heat}. It shows that, with low currents, the data scatter increases, even more than expected. In Table I, a small correction accounts for the thermometer self-heating and for a calibration repetition at the end of the measurements.

\begin{figure}
\centering
\includegraphics[width=75mm]{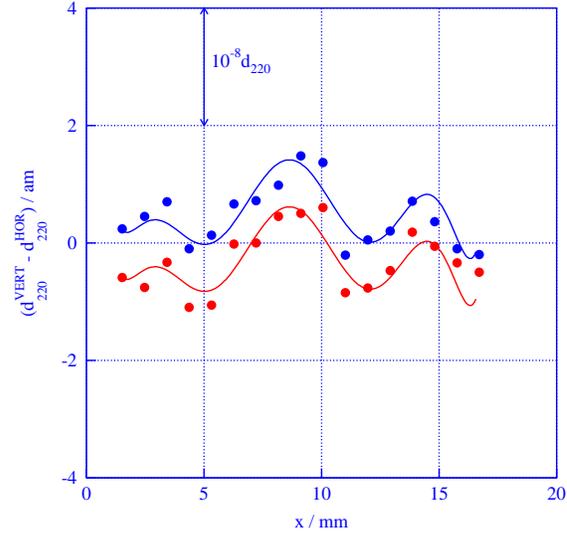}
\caption{Difference between the $\dd$ values in the vertical and horizontal pairs of the quadrant detector; measurements carried out on 2008-08-06 and 2008-09-16 are in red, those carried out on 2008-10-15 (with a replaced the pointing mirror) in blue. Solid lines are polynomial approximations of the data.}\label{astigmatism}
\end{figure}

The anomalies already mentioned in \cite{d220-OEXP} were further investigated. As shown in Fig.\ \ref{xint}, the analyzer displacement is sensed at four points, located on a circle of 1.5 mm diameter and spaced by a 90$^\circ$ angle. The displacements at the diametrically opposed points are made identical to within picometers by electronic control of the analyzer pitch and yaw rotations. In the absence of aberrations, that is, with perfectly flat or spherical wavefronts, also the displacements in any pair of contiguous points (separated by a 90$^\circ$ angle) are identical, though nothing prescribes them to be so. In other words, the four points of the analyzer front mirror are always expected to lie in a plane. The difference between the displacements of the vertical and the horizontal pairs of points is shown in Fig.\ \ref{astigmatism} in terms of its effect on the measured $\dd$ value; it gives clues about aberrations in the optical interferometer. The two sets of data refer to different configurations of the optical interferometer: the mirror sending the laser beam to the interferometer was replaced before the 2008-09-16 measurements were carried out. The data bias indicates astigmatism: the wavefront curvature in the horizontal and the vertical planes are different and so are the relevant corrections. The undulation indicates diffracting elements (dust particles, scratches, pits, and digs) in the optical interferometer. This anomaly undermines the confidence in the correction for diffraction and its uncertainty. However, a theoretical investigation \cite{fourier} suggests that the deviation of the effective wavelength of the integrated interference pattern with respect to the wavelength of a plane wave is proportional to the beam divergence, no matter what the angular spectrum of the laser beam might be \cite{goodman}. To confirm this, additional experimental investigations are under way.

\begin{figure}
\centering
\includegraphics[width=75mm]{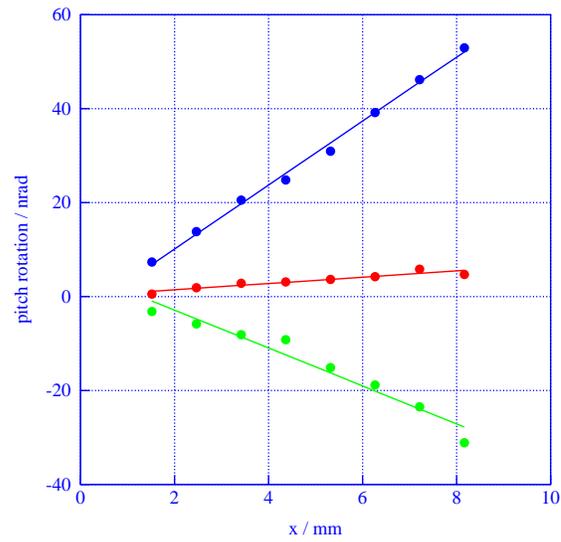}
\caption{Analyzer pitch rotation with imperfectly aligned laser beam. The nominal incidence angles of the beam on the analyzer front mirror are 90$^\circ$ (red), $90^\circ + 7$ arcseconds (green), and $90^\circ - 7$ aecseconds (blue).}\label{rho1}
\end{figure}

\begin{figure}
\centering
\includegraphics[width=75mm]{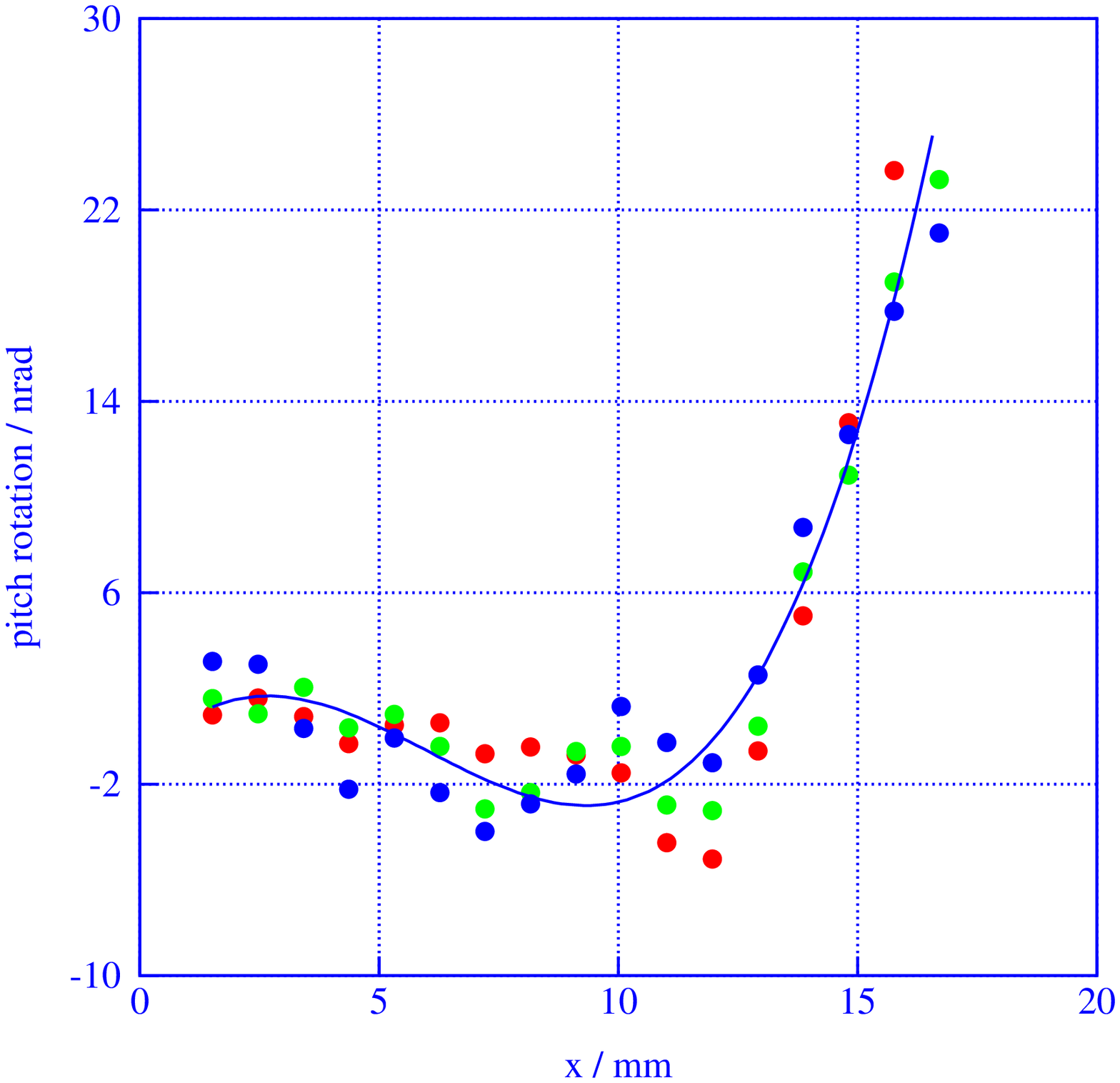}
\caption{Lattice plane tilt. Measurements were carried out on 2008-08-06 (red), 2008-09-16 (green), and 2008-10-15 (blue). A linear trend has been removed.}\label{rho2}
\end{figure}

Another anomaly is a difference between the x-ray and optical measurements of the analyzer pitch. As shown in Fig.\ \ref{xint}, the analyzer pitch and yaw rotations are measured and nullified by recording and by processing the optical interference pattern with a quadrant detector. To monitor the Abbe's error (difference between the displacements sensed by the x-ray and optical interferometers caused by rotations and offsets between the measure points), also the x-ray interference pattern was recorded and processed by means of a multianode photomultiplier with a vertical pile of NaI(Tl) scintillator crystals. In addition, the virtual anode having a zero vertical offset was identified and all the lattice spacing measurements were related to it. If the laser beam is not orthogonal to the analyzer front mirror, when the analyzer is moved, the reflected beam steps aside proportionally to the displacement. Owing to the wavefront curvature, this lateral drift originates an interference pattern imitating an analyzer rotation; this effect was discussed in \cite{fresnel}. Advantage was taken of the centimetre displacement capabilities to carry out more sensitive tests of this phenomenon. The laser beam was deviated, in the vertical plane, by $\pm 7$ arcseconds from the nominal 90$^\circ$ incidence and the analyzer pitch rotation was monitored by surveying the vertical gradient of the $\dd$ measurements. If the analyzer pitching is really null and the crystal lattice is really perfect, no gradient is expected. The integral of the observed gradient over the crossed lattice planes, expressed in terms of the equivalent analyzer pitching, is shown in Fig. \ref{rho1}. As expected, a non-orthogonal incidence causes the servo-control to detect and to compensate for a non-existent rotation.

Figure \ref{rho2} shows the results of the same analysis made on the data collected in the surveys in Fig.\ \ref{map}, which data extend to the whole crystal. Since the laser beam is now orthogonal to the analyzer front mirror, the observed rotation could be a clue to lattice imperfections. However, since the data derive from a comparison between x-ray and optical measurements, there is no really fair way to charge one device alone with the observed difference. A geometrical imperfection of the analyzer could explain the large lattice plane rotation observed at the crystal edge: the topographic image of the crystal surface is impressed in the phase of the spatial pattern of the x-ray fringes, and it is translated into an apparent lattice deformations.

\begin{figure}
\centering
\includegraphics[width=75mm]{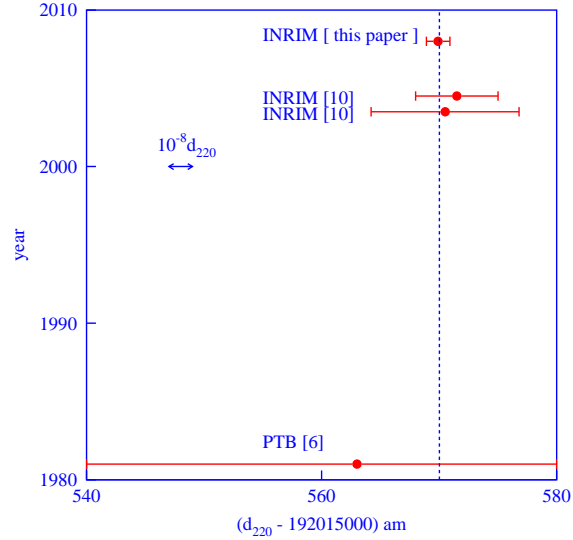}
\caption{Measurement values of the mean (220) lattice plane spacing in the WASO 4.2A crystal. The dash line is the weighed mean of the data.}\label{d220}
\end{figure}

\section{Conclusions}
A summary of the available measurement values of the WASO 4.2A lattice spacing is shown in Fig.\ \ref{d220} \cite{d220-PTB,waso42a-inrim}; the value here given confirms the shift of INRIM's results with respect to that of PTB. A comparison can also be made between the MO*4 and the WASO 4.2A lattice spacings obtained via absolute and relative measurements. This comparison is particularly meaningful because the MO*4 interferometer was sent to the PTB and directly compared against the WASO 4.2A one. In the absolute case, by using \cite{d220-OEXP}
\begin{equation}\label{d0}
 \dd({\rm MO*4}) = 192.0155508(12)  \textrm{ pm}
\end{equation}
and (\ref{d220value}), one obtains
\begin{equation}\label{d1}
 \dd({\rm MO*4}) - \dd({\rm WASO 4.2A}) = -18.3(1.6) \textrm{ am}.
\end{equation}
In the relative case, by averaging the data of Table 3 in \cite{Martin}, one obtains
\begin{equation}\label{d2}
 \dd({\rm MO*4}) - \dd({\rm WASO 4.2A}) = -18.9(1.9) \textrm{ am}.
% (102 +/- 13)E-9 and (95 +/- 14)E-9
\end{equation}

\ack
This work received funds from the European Community's Seventh  Framework Programme ERA-NET Plus -- grant 217257, from the Regione Piemonte -- grant D64, and from the Compagnia di San Paolo.

\end{document}